\documentclass[draft]{agujournal2018}
\usepackage{apacite}
\usepackage{url} 
\usepackage{lineno}

\journalname{JGR: Space Physics}

\begin{document}

\title{Studying dawn-dusk asymmetries of Mercury's magnetotail using MHD-EPIC simulations}

 \authors{Yuxi Chen\affil{1},
 {G\'abor T\'oth} \affil{1}, 
  Xianzhe Jia\affil{1},
  James A. Slavin\affil{1},          
  Weijie Sun\affil{1},
  Stefano Markidis\affil{2},
  Tamas I. Gombosi\affil{1},
  Jim M. Raines\affil{1}}

\affiliation{1}{Department of Climate and Space Sciences and Engineering, University of Michigan, , Ann Arbor, MI, USA}

\affiliation{2}{KTH, Stockholm, Sweden.} 


\correspondingauthor{Yuxi Chen}{yuxichen@umich.edu}

\begin{keypoints}
\item Dawnside current sheet is thicker than the duskside
\item Magnetic reconnection prefers dawnside under strong IMF driving
\item Almost all dipolarization events, high-speed proton and electron planetward flows concentrate on the dawnside
\end{keypoints}

\begin{abstract}
MESSENGER has observed a lot of dawn-dusk asymmetries in Mercury's magnetotail, such as the asymmetries of 
the cross-tail current sheet thickness and the occurrence of flux ropes,  dipolarization events and energetic electron injections. 
In order to obtain a global pictures of Mercury's magnetotail dynamics and the relationship between these asymmetries,
we perform global simulations with the magnetohydrodynamics with embedded particle-in-cell (MHD-EPIC) model, where Mercury's 
magnetotail region is covered by a PIC code. Our simulations show that the dawnside current sheet is thicker, the 
plasma density is larger, and the electron pressure is higher than the duskside. Under a strong IMF driver, the 
simulated reconnection sites prefer the dawnside. We also found the dipolarization events and 
the planetward electron jets are moving dawnward while they are moving towards the planet, so that
almost all dipolarization events and high-speed plasma flows concentrate in the dawn sector. 
The simulation results are consistent with MESSENGER observations. 
\end{abstract}

\section{Introduction}
MESSENGER has provided plenty of valuable information about Mercury's magnetosphere in the last decade, which have improved our understanding of the dynamics in the Mercury's magnetosphere. For examples, observations from MESSENGER have shown that the magnetospheric substorms at Mercury exhibit similar global magnetopheric configurations as the substorms at Earth, but in a time scale of 2 to 3 minutes, which is much shorter than the 2 to 3 hours of Earth's substorm \citep{Slavin:2010,Sun:2015}. MESSENGER has also observed magnetic structures that are closely related to magnetic reconnection, such as the flux transfer events near the magnetopause \citep{Slavin:2009,Slavin:2012b}, flux ropes or dipolarization fronts in the plasma sheet \citep{Slavin:2012,DiBraccio:2015,Sun:2015,Sun:2016}. These structures are similar to those in Earth's magnetosphere. However, at the same time, MESSENGER also found that several features are different from those of Earth. One of the most prominent puzzles raised by MESSENGER observations is the dawn-dusk asymmetry of Mercury's magnetotail.

Analyses of the MESSENGER data show that the energetic electrons or X-ray induced by energetic electrons on the nightside were more frequently observed in the postmidnight region, i.e., the dawnside, than in the premidnight region, i.e., the duskside \citep{Lindsay:2016,Baker:2016, Ho:2016, Dewey:2017}. The dawnward drifting of the electrons may explain the energetic electrons dawn-dusk asymmetry \citep{Lindsay:2016}. However, the study of magnetic reconnection related magnetic structures, which are flux ropes and dipolarization fronts, in the near-Mercury-neutral-line region showed both structures are also more frequently observed on the dawnside than on the duskside, which suggests the magnetic reconnection may prefer to happen on the dawnside and therefore created more energetic electrons in the postmidnight region than in the premidnight region [\citep{Sun:2016} see also, \citep{Sun:2017, Smith:2017}]. The dawnside magnetic reconnection preferentially occurrence in Mercury's plasma sheet is different from the observations in Earth's magnetosphere, where the magnetic reconnection related dynamic processes, such as the flux ropes \citep{Imber:2011} and dipolarization fronts \citep{Liu:2013b}, prefer the duskside plasma sheet. In addition, \citet{Poh:2017b} found Mercury's magnetotail current sheet is thicker on the dawnside than the duskside, and it is believed that it is easier to trigger magnetic reconnection in a thinner current sheet. The relationship between the current sheet thickness and the reconnection products observations still needs to be explored. It has also been observed that there are more heavy ions ($Na^{+}$ and $O^+$) on the duskside plasma sheet than in the dawnside plasma sheet \citep{Raines:2013, Gershman:2014}. The role of the heavy ions in the magnetic reconnection is still largely unknown.

Since the satellite observations usually localize to a small region of the whole magnetosphere at a given time, it is difficult to recover the timing sequence and the global picture of the magnetospheric dynamics from the localized data alone. Numerical models, especially global models, can provide  unique insight into these problems. \citet{Lin:2014}, \citet{Lu:2016} and \citet{Lu:2018} have used a global hybrid model and a local PIC model to study the dawn-dusk asymmetry of Earth's magnetosphere. They found that the Hall effect transports the current sheet plasma and the magnetic flux from the dusk sector to the dawn sector. The transportation reduces duskside current sheet thickness, thus reconnection is easier to be triggered on the duskside. This explanation may work for Earth, but there are some difficulties to adopt it for Mercury. Mercury's current sheet is thinner \citep{Poh:2017b} on the duskside, which is similar to the Earth and might be explained by the Hall effect. However, Mercury's reconnection products prefer the dawn sector. Recently, \citet{Liu:2019} used box PIC simulations to study the magnetic reconnection preference for a thin current sheet that is embedded into a thick current sheet, and they found there is an inactive region on the ion drifting side, and therefore the reconnection prefers the electron drifting side, which might be applicable at Mercury.

A global numerical model of Mercury's magnetosphere is needed to solve these puzzles. 
Several numerical models have been used to study Mercury's magnetosphere in the past decades. 
BATS-R-US was the first MHD model applied for 3D global simulations of Mercury's 
magnetosphere \citep{Kabin:2000, Kabin:2008}. \citet{Jia:2015, Jia:2019} developed
the resistive body capability for BATS-R-US and studied how the induction effect that is arising from the 
conducting core affects the magnetospheric global response to the varying solar wind conditions. 
Multi-fluid MHD models that treat heavy ions as a separate fluid have been used for Mercury's 
magnetosphere simulations \citep{Kidder:2008}. Since the kinetic scales of Mercury's magnetospheric 
plasma can be comparable to Mercury's radius, kinetic effects may play an important 
role in Mercury's magnetosphere. To incorporate kinetic physics, hybrid 
models \citep{Kallio:2003, Wang:2010, Muller:2012, Travnicek:2010}, 
which treat the electrons as a massless charged fluid and model the ions as particles, 
test particle models, which trace the particle trajectories with a global electromagnetic field 
obtained from either a global numerical model \citep{Schriver:2011, Seki:2013} or an 
analytic model \citep{Delcourt:2003, Delcourt:2013}, and particle-in-cell 
models \citep{Schriver:2017} have been applied to study Mercury's magnetosphere. 
Due to the limitations of the physics capabilities or the grid resolutions of these models, the 
dawn-dusk asymmetries of Mercury's magnetotail have not been studied in detail. 

The MHD with embedded PIC (MHD-EPIC) model \citep{Daldorff:2014} makes it feasible 
to study Mercury's magnetotail dynamics with a realistic configuration. We use a PIC code
to cover Mercury's inner tail, and the rest of the domain is handled by the MHD model BATS-R-US. 
The details of the numerical model are discussed in section~\ref{section:model}. 
Section~\ref{section:observation} provides the MESSENGER data that is used to 
compare with simulations in the later sections. The simulation results are presented 
and discussed in section~\ref{section:results} and section~\ref{section:discussion}. 

\section{Numerical model}
\label{section:model}
The MHD-EPIC model has been successfully applied to investigate the interaction between 
the Jovian wind and Ganymede's magnetosphere \citep{Toth:2016, Zhou:2019}, Martian magnetotail reconnection \citep{Ma:2018} 
and Earth's dayside reconnection \citep{Chen:2017, Toth:2017}. The MHD-EPIC model two-way couples the Hall
MHD model BATS-R-US \citep{Powell:1999, Toth:2008}  and the semi-implicit particle-in-cell code 
iPIC3D \citep{Markidis:2010} through the Space Weather Modeling Framework 
(SWMF) \citep{Toth:2005swmf, Toth:2012swmf}. Recently, \citet{Chen:2019} has developed the Gauss's Law 
satisfying Energy Conserving Semi-Implicit Method (GL-ECSIM), an improved version of the ECSIM \citep{Lapenta:2017a}, and implemented it into 
the iPIC3D code. This new PIC algorithm is used for all the MHD-EPIC simulations presented here. 

For the MHD-EPIC simulations of Mercury's magnetosphere, we run the fluid code BATS-R-US first 
to reach a steady state, then we change to the time-accurate mode \citep{Toth:2012swmf} 
and couple the fluid model with the PIC code. Hall-MHD equations are solved by the fluid model for 
both MHD-EPIC simulations and pure Hall-MHD simulations. The simulation 
setup for both BATS-R-US and PIC are described in the following subsections. 

\subsection{Global MHD model: BATS-R-US}
Following the work of \citet{Jia:2015}, a resistive body with finite conductivity layer is 
used to represent the interior structure of Mercury: the region within $r<0.8\,R_M$ is the 
highly conducting core, and the layer between $0.8\,R_M$ and $1\,R_M$ with finite conductivity 
represents the mantle. The conductivity inside the mantle is set to be $\sim 10^{-7}\,S/m$. We refer 
to \citet{Jia:2015} for more details about the conductivity profile. 

The Hall effect
and the electron pressure gradient term are also included in our generalized Ohm's law:
\begin{equation} 
\mathbf{E} = -\mathbf{u}\times \mathbf{B} + \frac{\mathbf{J}\times \mathbf{B}}{q_e n_e}-
\frac{\nabla p_e}{q_e n_e} + \eta \mathbf{J}
\end{equation}\label{eq:Ohm1}
where $q_e$, $n_e$ and $p_e$ are the unsigned electron charge, electron number density (obtained from charge neutrality) and 
electron pressure, respectively. $\eta$ represents the resistivity, which is the 
inverse of the conductivity and $\mathbf{J}=\nabla\times\mathbf{B}/\mu_0$ is the current density. The electron pressure is calculated from a separate equation:
\begin{equation} 
\frac{\partial p_e}{\partial t} + \nabla \cdot (p_e \mathbf{u}_e) = (\gamma -1) (-p_e \nabla \cdot \mathbf{u}_e )
\end{equation}\label{eq:pe}
where $\gamma = 5/3$ is the adiabatic index, and $\mathbf{u_e} = \mathbf{u} - \mathbf{J}/(q_en_e)$ 
is the electron velocity. In summary, the resistive Hall MHD equations with a separate electron 
pressure equation are solved in our MHD model. 

Inside the mantle region ($0.8\,R_M<r<1\,R_M$), 
there is no plasma flow, but the magnetic field still changes due to the finite conductivity. 
Only the reduced Faraday's law is solved inside the mantle: 
\begin{equation}
\frac{\partial \mathbf{B}}{\partial t} = -\nabla \times (\eta \mathbf{J}).
\end{equation}
Outside the planet surface, 
the whole set of MHD equations are solved. Since both the Hall term and the resistivity term are stiff, 
a semi-implicit scheme \citep{Toth:2012swmf} is used to speed up the simulations: the equations 
excluding the stiff terms are solved explicitly first, then the stiff terms are solved by an implicit solver. 

The simulations are performed in the Mercury solar orbital (MSO) coordinates, where the X-axis is pointing to 
the Sun from Mercury, the Z-axis is parallel to Mercury's rotation axis, and the Y-axis completes a 
right-handed coordinate system. 
The whole simulation domain is a brick of $-64\,R_M<x<8\,R_M$ and $-32\,R_M<y,z<32\,R_M$ cut out from a spherical grid. The center 
of Mercury coincides with the origin of the coordinates. A dipole field with strength of $200\,\mathrm{nT}$ \citep{Anderson:2011}
at the magnetic equator is used. The dipole axis is aligned with the Z-axis but the dipole center is shifted 
northward by $0.2\,R_M$. A stretched locally refined spherical grid is used. The tail region is refined so 
that the cell size is about $0.025\, R_M$ near $x=-2.5\, R_M$. From our simulations, the plasma density in the 
lobes is about $0.3\,\mathrm{amu/cm^3}$, and the corresponding proton inertial length is about $360\,\mathrm{km}$ 
or $0.15\,R_M$. The Hall effect can be well resolved because one inertial length is covered by $\sim 6$ cells. 
The inner boundary condition for the magnetic field is applied at the interface of the mantle and the conducting 
core, where $r=0.8\,R_M$ and the magnetic field is fixed due to the high conductivity. Since there is no plasma 
flow in the mantle, the inner boundary conditions for plasma density, velocity and pressure are applied 
on the planet surface $r=1\,R_M$. 
A zero gradient boundary condition is applied to plasma density and pressure. The boundary condition for velocity 
is designed so that the plasma can be absorbed by the surface, and the surface is not an important source of plasma. 
For the inflow, a zero gradient boundary condition is applied to all velocity components. For the outflow, the radial 
velocity component is set to be zero at the boundary and a zero gradient boundary condition is applied to the 
tangential components. The plasma may flow around or flow into the surface, but it would not have a 
significant outflow component. 

\subsection{PIC model}
\begin{figure}
 \includegraphics[width=\textwidth]{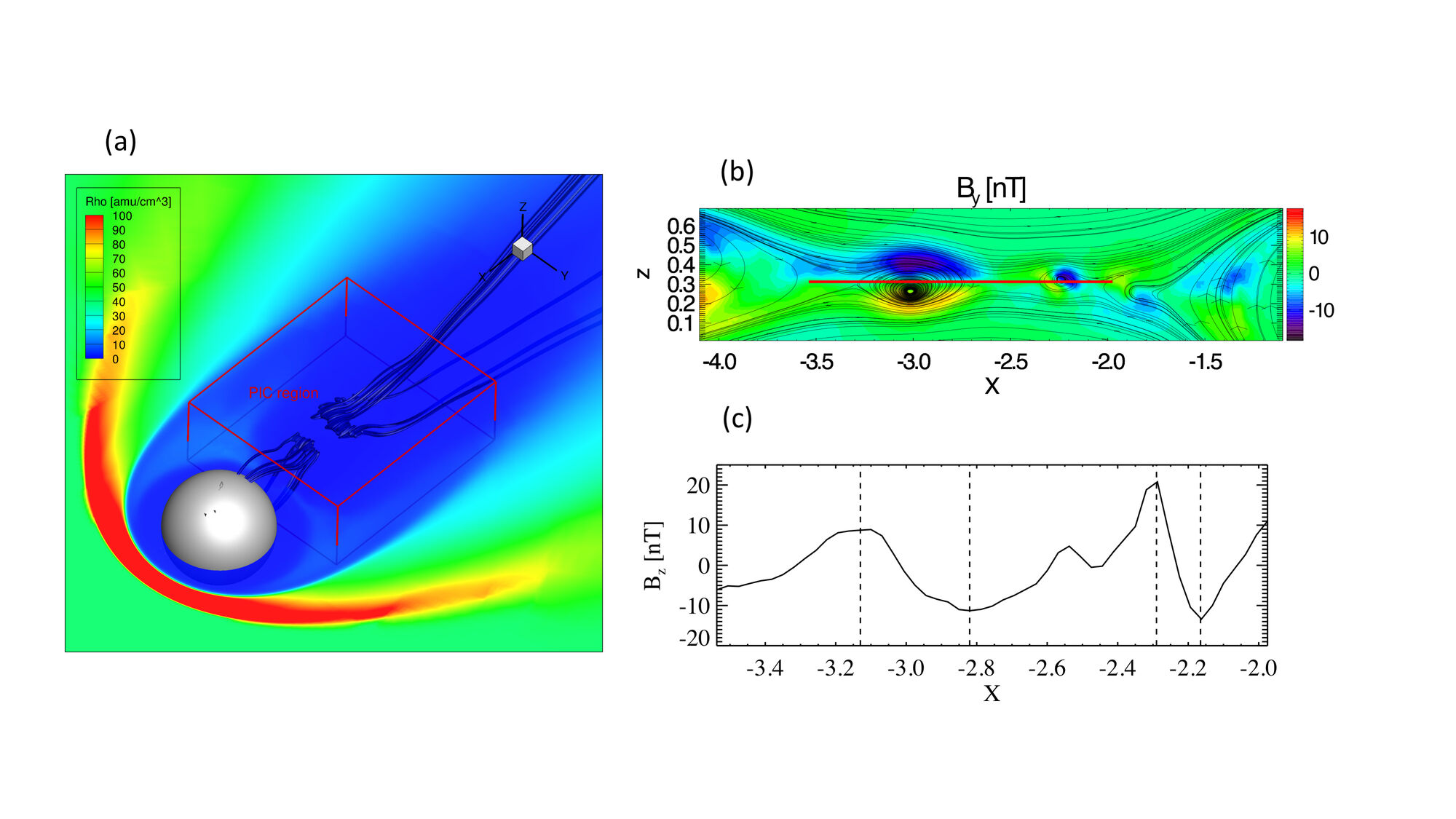}
 \caption{(a) The global structure of Mercury's magnetosphere at $t = 300\,s$ from the simulation MHD-EPIC-A. 
 The mass density in the equatorial plane and the magnetic field lines of two flux ropes are shown. 
 The red box is the region covered by the PIC code. It covers the whole tail region where magnetic 
 reconnection may happen. In the Y-direction, the PIC region is close to but has not reached the 
 magnetopause. (b) The Hall magnetic field $B_y$ and magnetic field lines at $y=0$. (c) The $B_z$ component along the red line in (b). 
 }
 \label{fig:pic_region}
 \end{figure}

The Gauss's Law satisfying Energy Conserving Semi-Implicit Method (GL-ECSIM) \citep{Chen:2019} is used in the PIC region. MESSENGER observations suggest that the average near-Mercury neutral line (NMNL) is at around $x=-3\,R_M$ \citep{Slavin:2009,Poh:2017a}. To study Mercury's magnetotail reconnection, the tail region $-5.1\,R_M<x<-1.1\,R_M$, $-1.75\,R_M<y<1.75\,R_M$ and $-0.5\,R_M<z<1.5\,R_M$ is covered by the uniform Cartesian mesh of the PIC code (see Figure~\ref{fig:pic_region}(a)). The cell size is $1/32\,R_M$ in all directions. 64 macro-particles per species per cell are used. In order to reduce the computational cost, an artificially reduced proton-electron mass ratio of $m_p/m_e=100$ is set. The cell size is $\sim 1/5 $ of the proton inertial length or twice of the plasma skin depth.  The time step is $2.5\times10^{-3}$ s, the maximum electron thermal speed is about $8\times 10^3$ km/s, and the cell size is $1/32\,R_M$, so that the corresponding CFL number (the ratio of the time step to the cell crossing time by electrons) is about 0.25, which satisfies the `accuracy condition' of the semi-implicit PIC methods \citep{Markidis:2010}. 

 \section{MESSENGER observations in the nightside plasma sheet}
 \label{section:observation}
 This section provides observations of the proton properties and dipolarization fronts in Mercury's nightside plasma sheet from MESSENGER \citep{Solomon:2007}. The proton measurements are provided by the Fast Imaging Plasma Spectrometer (FIPS) \citep{Andrews:2007} and the magnetic field measurements are provided by the magnetometer \citep{Anderson:2007}. FIPS could measure ions in an effective field of view of $\sim 1.15 \pi$~sr with an energy range from $\sim 46$ eV/e to $\sim 13.7$ keV/e with a time resolution of $\sim 10$~s. The magnetic field data are provided with a time resolution of 20 vectors per second and are under Mercury solar magnetospheric coordinates (MSM). In the MSM, the $X_{MSM}$ is sunward, the $Z_{MSM}$ is northward and parallel to the dipole axis, and the $Y_{MSM}$ completes the right-handed system. The MSM coordinates and MSO coordinates are parallel with each other, but the MSO origin is the center of Mercury and the MSM origin centers 
 on the Mercury dipole. The spacecraft position data are provided to be in the same time resolution as the magnetic field measurements, but they are aberrated according to the solar wind velocity and Mercury's orbital motion to make the $X'_{MSM}$ antiparallel to the solar wind. The aberration changes the positions in the $X_{MSM}-Y_{MSM}$ plane, but does not change $Z_{MSM}$.
 
 \begin{figure}
 \includegraphics[width=1\textwidth, trim=0cm 6cm 0cm 6cm]{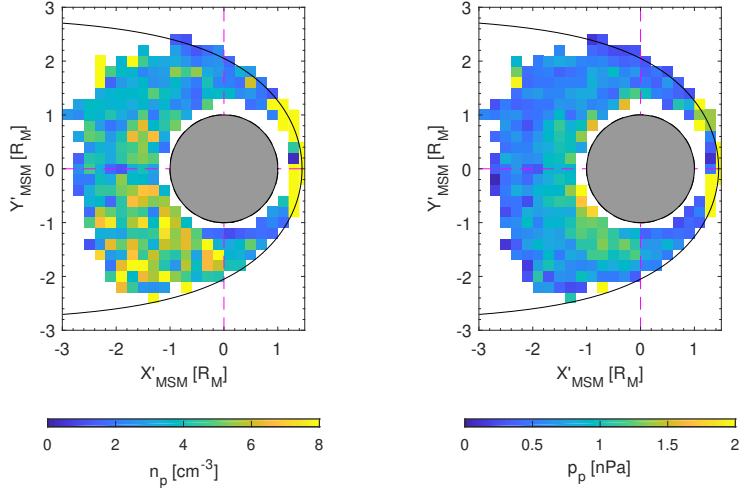}
 \caption{The MESSENGER observed proton density (left) and proton pressure (right) around the magnetic equator ($|Z_{MSM}| < 0.2 R_{M}$). This figure shows the one minute proton moments derived from FIPS during the entire MESSENGER orbits around the Mercury (from 17 March 2011 to 30 April 2015). The black curve on both figures is the average location of the magnetopause \citep{Winslow:2013}. The number of events in each bin is required to be larger than 3. The size of bin is $0.2 \ R_{M} \times 0.2 \ R_{M}$. The colors indicate the intensity of density (left) and pressure (right), respectively.}
 \label{fig:observation_plasma}
 \end{figure}
 
 \subsection{Proton properties}
 Proton density and pressure shown in Figure~\ref{fig:observation_plasma} were derived from one minute average distributions of protons under the assumption that they are isotropic and stationary Maxwellian distributions \citep{Raines:2011, Raines:2013,Gershman:2013}. The proton moments derived from this method were applied in several studies on the plasma sheet dynamics \citep{Raines:2011,Gershman:2014,Sun:2017,Sun:2018,Poh:2018}. The proton density distribution (left figure in Figure~\ref{fig:observation_plasma}) shows clear dawn-dusk asymmetry with proton densities higher on the dawnside ($\sim 6$ to 8~amu/cc) than on the duskside ($\sim 2$ to 4~amu/cc). The dawn-dusk asymmetry of proton pressure (right figure in Figure~\ref{fig:observation_plasma}) is not that prominent as proton density. The proton pressure shows weak dawn-dusk asymmetry in the downtail region $(X'_{MSM} < - 1.3 \ R_{M})$ with proton pressure on the dawnside plasma sheet slightly higher than on the duskside. This dawn-dusk asymmetry becomes more prominent in the near tail region with ($X'_{MSM} \sim - 1 \ R_{M}$), where proton pressure was from 1.3 to 1.7~nPa on the dawnside plasma sheet and was from 0.6 to 1.3~nPa on the duskside plasma sheet.
 
 \citet{Korth:2014} showed the distribution of mean proton flux in the nightside plasma sheet of Mercury. In that study, the mean proton flux showed clear dawn-dusk asymmetry with the flux much higher on the dawnside than on the duskside, which is similar to the distribution of proton density in Figure~\ref{fig:observation_plasma}.

  \begin{figure}
 \includegraphics[width=1\textwidth, trim=0cm 6cm 0cm 6cm]{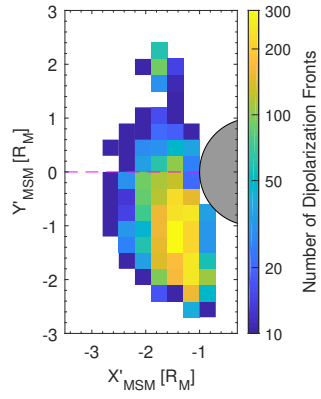}
 \caption{The spatial distribution of the dipolarization fronts observed by MESSENGER around the magnetic equator ($|Z_{MSM}| < 0.2 R_{M}$). The size of bin is $0.3 \ R_{M} \times 0.3 \ R_{M}$. The color indicates the number of dipolarization fronts in each bin. The number of dipolarization fronts in each bin is required to be at least 3.}
 \label{fig:observation_df}
 \end{figure}

 \subsection{Dipolarization fronts}
 Dipolarization front, also called reconnection front, is defined as the leading edge of planetward travelling plasma flow burst, which is highly correlated with the magnetic reconnection [e.g., \citep{Angelopoulos:2013}]. In previous studies at Mercury, \citet{Sun:2016} has shown clear dawn-dusk asymmetry of dipolarization fronts in the near-Mercury-neutral-line region with more dipolarization fronts on the dawnside plasma sheet than on the duskside plasma sheet. The following studies on the dipolarization fronts in the near-Mercury plasma sheet, proton energization and heating, energetic electrons and proton bulk flows have shown the similar dawn-dusk asymmetries \citep{Sun:2017,Dewey:2017,Dewey:2018}.

 
 Figure~\ref{fig:observation_df} shows the distribution of dipolarization fronts in Mercury's nightside plasma sheet. This figure contains the dipolarization fronts during the entire period MESSENGER orbited around Mercury. The dipolarization fronts were obtained according to the similar procedure as \citet{Sun:2016}. Since the dipolarization fronts were constrained in the regions with $Z_{MSM} < 0.2 \ R_{M}$ and MESSENGER orbits were evenly distributed in the dawn-dusk direction \citep{Sun:2016}, the occurrence rate of dipolarization fronts shows essentially the same structures as Figure~\ref{fig:observation_df}. In the downtail region $(X'_{MSM} < - 2 \ R_{M})$, the dipolarization fronts show dawn-dusk asymmetry with more events on the dawnside plasma sheet than on the duskside, which is similar to \citep{Sun:2016}. The dawn-dusk asymmetry becomes more prominent in the region closer to the planet (from $-2 \ R_{M}$ to $-1 \ R_{M}$).

\section{Simulation results}
\label{section:results}
We perform pure Hall-MHD and MHD-EPIC simulations with different upstream solar wind conditions. 
In order to avoid introducing dawn-dusk asymmetries from the solar wind, the Y-components of 
the IMF and the solar wind velocity are eliminated in all simulations. Since the Y-component 
of the velocity is zero, there is not need to apply aberration to the simulation results.  
The detailed solar wind parameters are shown 
in Table~\ref{tb:sw}. Compared to the parameters used by \citet{Jia:2015}, we use a proton
and electron temperature of 7.5~eV, which is half of the proton temperature of \citet{Jia:2015}. 
Since the total pressure of the solar wind is split between electrons and protons in this paper, 
the total plasma thermal pressure is still the same as \citet{Jia:2015}. The strength of the IMF in both
MHD-EPIC-A/Hall-A and MHD-EPIC-B/Hall-B is $|\mathbf{B}|=19.4$ nT, which are also the same as \citet{Jia:2015}. 
The plasma parameters for MHD-EPIC-A/Hall-A is typical at Mercury's ambient space environment.
The IMF configuration of MHD-EPIC-A/Hall-A is similar to a typical Parker spiral magnetic field, 
except that the $B_y$ component is set to be zero and a negative $B_z$ component is 
introduced to drive Mercury's magnetosphere. The IMF of MHD-EPIC-B/Hall-B purely consists of a
negative $B_z$ component with larger magnitude, which is a stronger driver than that of
MHD-EPIC-A/Hall-A. We run the MHD code first to reach a steady state, then we run the 
time-accurate MHD-EPIC or Hall-MHD for 300~s, which is about 2 to 3 Dungey cycles of Mercury's 
magnetosphere \citep{Slavin:2009}. It usually takes a numerical 
model a few Dungey cycles to settle down to a steady or quasi-steady state. 

In the following subsections, we introduce the global picture of the simulation results first. 
Then the dawn-dusk asymmetry is discussed based on the simulations. We will briefly compare
the MHD-EPIC simulations with the pure Hall-MHD simulations as well.  

\begin{table}[ht]                                                                                   
\caption{The solar wind parameters in MSO coordinates.}                    
\centering                                                                                          
\begin{tabular}{c c c c c c }                                                                       
\hline                                                                                              
 Simulation ID    & $\rho$ [amu/cc] & Temperature [eV]  & Velocity km/s  & IMF\,[nT] \\          
\hline                                                                                              
MHD-EPIC-A/Hall-A    & 40         & 7.5                 &  (-400, 0, 0)  & (-17.4, 0, -8.5)   \\                     
MHD-EPIC-B/Hall-B    & 40         & 7.5                 &  (-400, 0, 0)  & (0, 0, -19.4)   \\
\hline                                                                                              
\label{tb:sw}                                                                               
\end{tabular}                                                                                       
\end{table}                                                                                         
                                                                                                    

\subsection{Global picture}
The global structure of Mercury's magnetosphere at $t = 300\,s$ from the simulation MHD-EPIC-A
is shown in Figure~\ref{fig:pic_region}. The equatorial plane is colored by the plasma mass density. 
It happens to have two flux ropes at this moment. By checking the time series of 
the simulation results, it is easy to figure out that the flux rope far from the planet 
is moving tailward, and the one near Mercury is moving planetward. These flux ropes 
are produced by the PIC code, which covers most parts of the inner magnetotail. In the 
Y-direction, the PIC region is close to but has not reached the magnetopause. 
Figure~\ref{fig:pic_region} shows a typical state of the MHD-EPIC-A simulation. 
Magnetic reconnection happens around $x = -2.5\,R_M$, and produces tailward and planetward moving
flux ropes. 

A 2D cut at y = 0 is presented in Figure~\ref{fig:pic_region}(b) to show more details of 
these two flux ropes. The bipolar $B_y$ field is 
the remnants of the reconnection Hall magnetic field. There is no significant core field for 
either flux ropes at this moment due to the lack of IMF $B_y$, which may act as 
core field seed during the formation of a flux rope. Compared to a typical flux rope with a strong 
core field, these two flux ropes presented here are more like collections of O-lines. The tailward 
flux rope is about $1\,R_M$ long in the Y-direction, and the planetward one is about $0.5\,R_M$ long. 
The flux rope diameter measured by the $B_z$ field peak-to-peak distance in the X-direction is about
$0.3\,R_M$ (730 km) for the tailward one and $0.15\,R_M$ (360 km) for the planetward one (Figure~\ref{fig:pic_region}(c)). \cite{DiBraccio:2015} estimates the mean flux rope diameter to be 
$0.14\,R_M$ (345 km) by using the Alfven speed of 465~km/s times the time delay between 
MESSENGER detecting the two $B_z$ peaks. Our simulations suggest the typical ion jet velocity 
is about 1000~km/s (Figure~\ref{fig:cs_ux}). The mean diameter of the MESSENGER observed flux ropes
will be about $0.3\,R_M$ if 1000~km/s instead of 465~km/s is used in the estimation. In any case, 
the diameters of the two flux ropes in Figure~\ref{fig:pic_region} are similar to the MESSENGER observations.
Across the flux ropes, $B_z$ changes from 10~nT to -10~nT for the tailward one and from 20~nT to 
-15~nT for the planetward one. These $B_z$ peak-to-peak amplitudes are close to the average of MESSENGER
observation value of 20~nT \citep{DiBraccio:2015}. Inside the flux rope, the proton density is about 
1.5~amu/cc in the simulation, while the median observed density is 2.03~amu/cc \citep{DiBraccio:2015}.

The agreement of the flux rope properties between the MHD-EPIC-A simulation and MESSENGER observations 
demonstrates that our mode behaves reasonably well in capturing Mercury's magnetotail reconnection. 
In the following subsections, we will examine the dawn-dusk asymmetries of Mercury's tail.

\subsection{Tail current sheet thickness and plasma profile}

\begin{figure}
 \includegraphics[width=1.0\textwidth, trim=0cm 0cm 0cm 0cm]{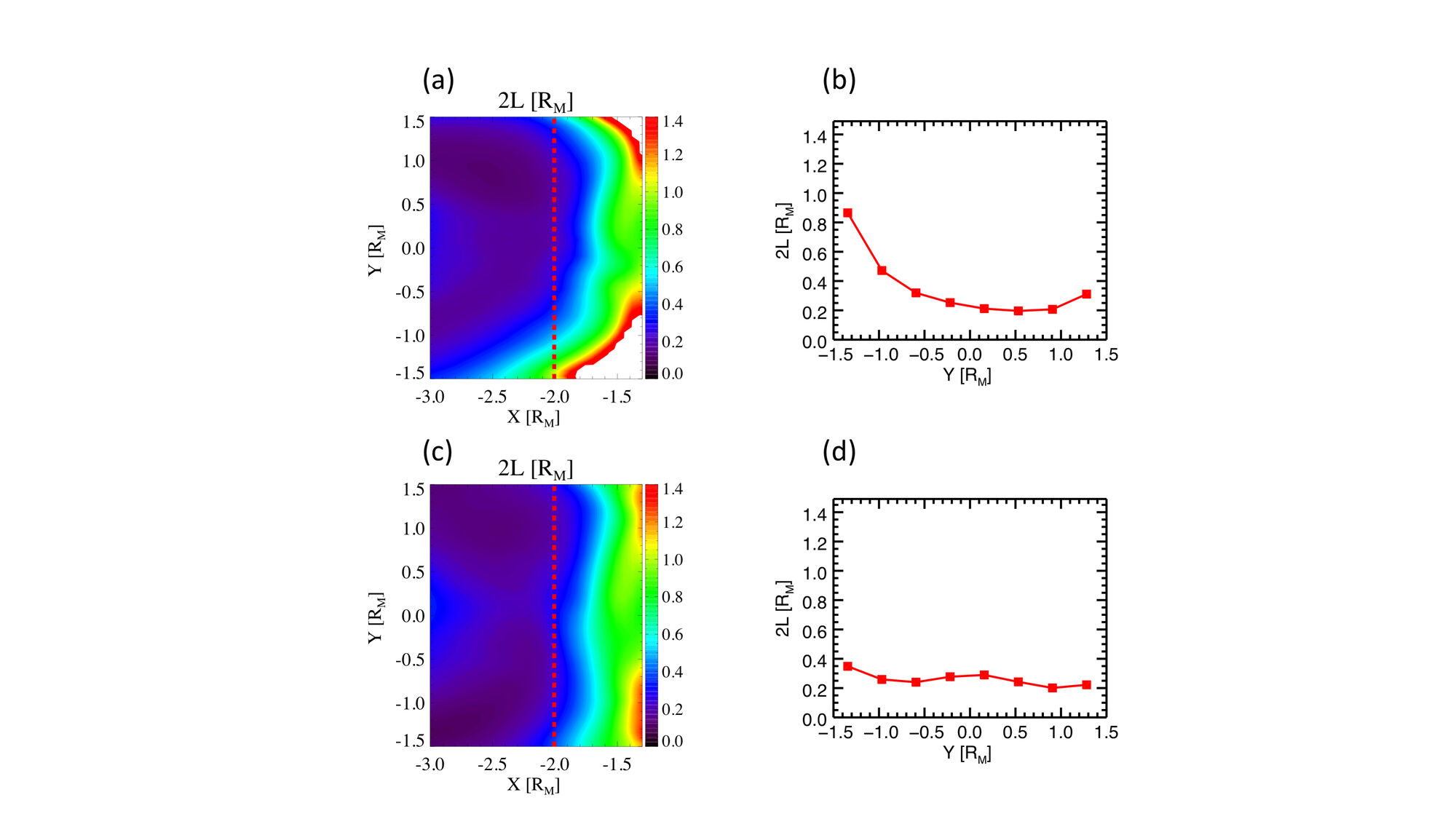}
 \caption{The time-averaged current sheet thickness of MHD-EPIC simulations. 
 (a) and (c) correspond to the MHD-EPIC-A and MHD-EPIC-B runs, respectively. 
 (b) and (d) are the thickness at $x=-2\,R_M$, which is marked by the red dashed lines in (a) and (c).}
 \label{fig:cs}
 \end{figure}
 
 \begin{figure}
 \includegraphics[width=1.0\textwidth, trim=0cm 0cm 0cm 0cm]{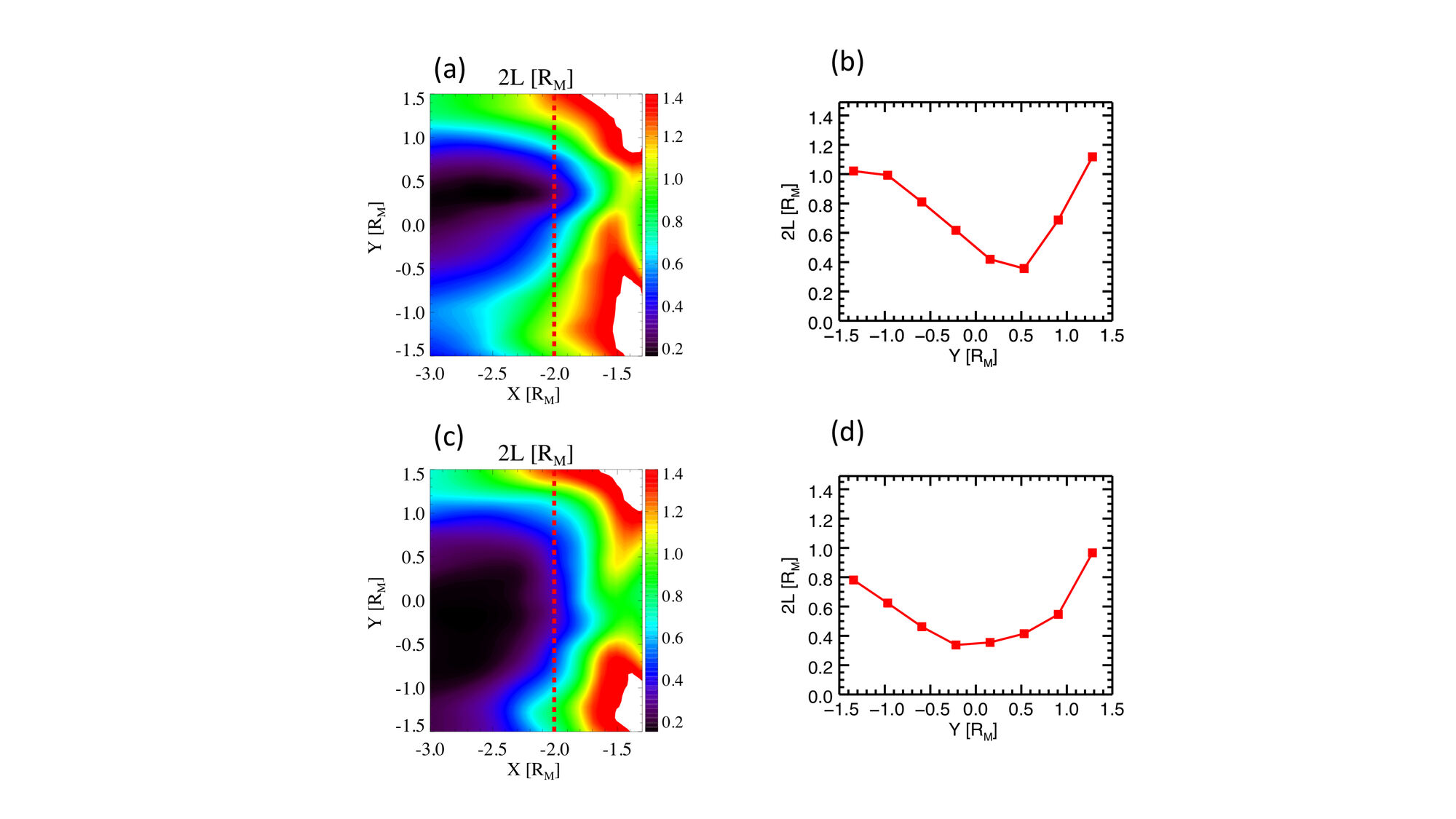}
 \caption{The current sheet thickness of Hall-A ((a) and (b)) and Hall-B ((c) and (d)) simulations. 
 (b) and (d) are plots of current sheet thickness at $x=-2\,R_M$.}
 \label{fig:cs_hall}
 \end{figure}
 
  \begin{figure}
 \includegraphics[width=1.0\textwidth, trim=0cm 0cm 0cm 0cm]{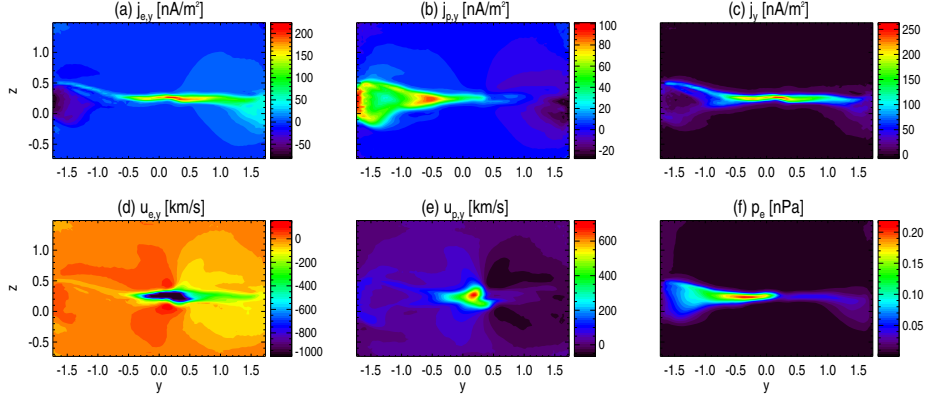}
 \caption{The 300-second average current sheet structure of MHD-EPIC-A at $x=-2\,R_M$. 
 (a), (b) and (c) are the electron current, proton current and total current in the 
 Y-direction, respectively. (d) is the electron velocity in the Y-direction. It 
 can be as fast as -4,000 km/s, and we make the color saturated at -1,000 km/s to 
 show more structures. (e) is the proton velocity in the Y-direction. (f) is the
 electron pressure.}
 \label{fig:current}
 \end{figure}

\citet{Poh:2017b} calculated the current sheet thickness from hundreds of MESSENGER crossings, and they found
the current sheet is thinner on the duskside (+Y) than the dawnside (-Y) on average. 
Using the same fitting method described by \citet{Poh:2017b}, we calculate the 
current sheet thickness from our simulations. The $B_x$ field along the Z-axis 
is fitted to a one-dimensional Harris current sheet model: 
\begin{equation} 
B_x (z) = B_0 \tanh \left(\frac{z - z_0}{L} \right).
\end{equation}\label{eq:cs}
The fitted current sheet thickness is 2L. 
The fitting is done every 2\,s, and its average over 300\,s is shown in Figure~\ref{fig:cs}. 

Figure~\ref{fig:cs}(a) shows that the center of the thin current sheet of MHD-EPIC-A 
is shifted to the dusk (+Y) direction. The proton density of the thin current sheet
around $x=-2\,R_M$ is about 0.4 amu/cc, and it can be as low as 0.02 amu/cc in the ambient lobe. The
proton inertial length $d_i$ of a density of 0.4 amu/cc is $0.15\,R_M$, which is the same order as 
the current sheet thickness in Figure~\ref{fig:cs}. A cut of thickness at $x=-2\,R_M$ 
is presented in Figure~\ref{fig:cs}(b). It is clear to see that the current sheet is thicker 
on the dawnside (-Y) than the duskside (+Y), which is consistent with the profile obtained 
from MESSENGER data. Figure 4(b) of \citet{Poh:2017b} shows the current sheet thickness from 
hundreds of current sheet crossings. For this MESSENGER plot, the corresponding solar wind conditions 
are unknown and may vary a lot, and it contains current sheet crossings from $x=-3.0\,R_M$ to $x=-1.1\,R_M$, 
so the thickness may vary from $0.1\,R_M$ to $1\,R_M$ even for the same Y coordinate. 
But the mean current sheet thickness is probably able to represent the status under a typical 
solar wind condition. In the observation plot, the thinnest average current sheet is
about $0.3\,R_M$, and it increases to about $0.7\,R_M$ on the dawnside and $0.5\,R_M$ on the duskside.
Since $x=-2\,R_M$ is roughly the middle point of the MESSENGER crossings distribution in the 
X-direction, we plot the current sheet thickness at $x=-2\,R_M$ in Figure~\ref{fig:cs}(b). 
The current sheet can be as thin as $0.2\,R_M$, and it increases to $0.8\,R_M$ 
at $y=-1.5\,R_M$ and $0.3\,R_M$ at $y=1.5\,R_M$. The MHD-EPIC-A simulation 
current sheet is slightly thinner than the observations around
midnight and in the dusk sector. Considering the large variance in the MESSENGER 
data (Figure 4(b) of \citet{Poh:2017b}), the simulation agrees with observations very well. 

The current sheet thickness for MHD-EPIC-B, which is driven by $B_z=-19.4$~nT IMF, is presented in 
Figure~\ref{fig:cs}(c) and (d). The current sheet that is far away from the midnight becomes 
thinner than in the MHD-EPIC-A simulation, because the stronger dayside magnetic reconnection transports
more magnetic flux to the tail to produce higher magnetic pressure. The thickness becomes less 
asymmetric than MHD-EPIC-A, even though the dawnside current sheet is still slightly thicker 
than the duskside. The bump near the midnight is probably
produced by the thick current sheet of the reconnection exhaust. 

We repeat the same analysis of the current sheet thickness for the two Hall-MHD 
simulations using the same input parameters as those in the MHD-EPIC simulations. 
The results are shown in Figure~\ref{fig:cs_hall} for comparison. The current sheet 
thickness at X = -2 $R_M$ in the Hall MHD simulations is significantly larger 
compared to the MHD-EPIC simulation results and the MESSENGER observations. It can 
be seen that the current sheet thickness is not symmetric around midnight in the 
Hall-MHD simulations, either, and the thinnest part of the tail current sheet is 
displaced towards dusk (+Y), which is similar to that seen in the MHD-EPIC simulations. 
These results together suggest that the asymmetry is likely to be related to the Hall effect.

The cross-tail current density of MHD-EPIC-A at $x=-2\,R_M$ is presented in 
Figure~\ref{fig:current}. The duskside (+Y) electron current density is larger 
than the dawnside (-Y), but the proton current density is larger on the dawnside (-Y).
The maximum current density of $j_y \approx 200\,$nA/m$^2$ arises around midnight, 
and it reduces to less than $100\,$nA/m$^2$ on the two flanks. The thin current sheet 
extends farther in the dusk sector (+Y) than the dawn sector. The spatial variation of 
the total current density $j_y$ presented here is consistent with MESSENGER observations
(Figure~4(c) of \citet{Poh:2017b}). 

\begin{figure}
 \includegraphics[width=1.0\textwidth, trim=0cm 0cm 0cm 0cm]{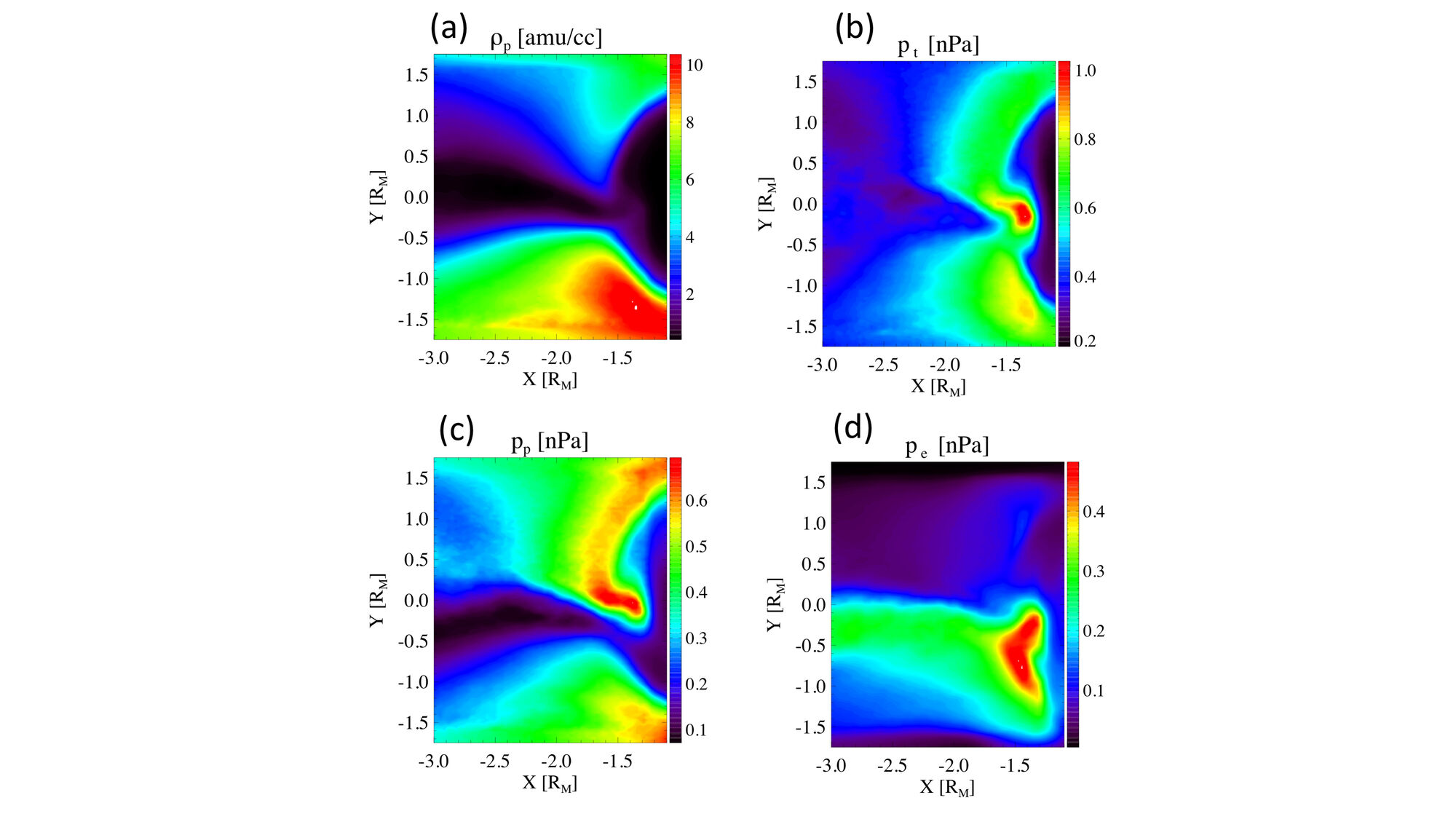}
 \caption{The time-averaged plasma profiles from the PIC output on the current sheet surface for MHD-EPIC-A: proton density (a), total pressure (b), proton pressure (c) and electron pressure (d).}
 \label{fig:cs_av_epic_a}
 \end{figure}

\begin{figure}
 \includegraphics[width=1.0\textwidth, trim=0cm 0cm 0cm 0cm]{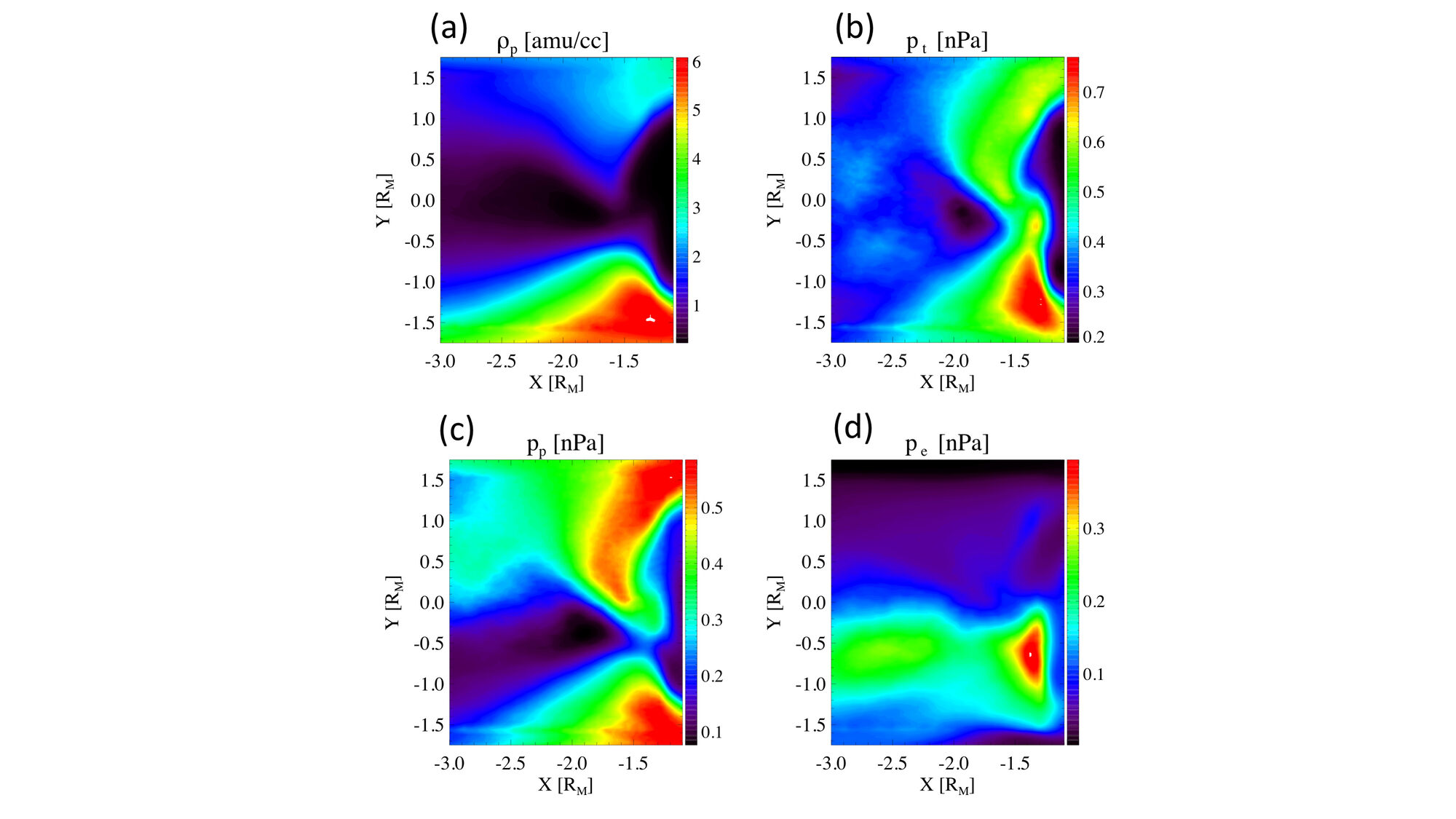}
 \caption{The time-averaged plasma profiles from the PIC outputs on the current sheet surface for MHD-EPIC-B.}
 \label{fig:cs_av_epic_b}
 \end{figure}

\begin{figure}
 \includegraphics[width=1.0\textwidth, trim=0cm 0cm 0cm 0cm]{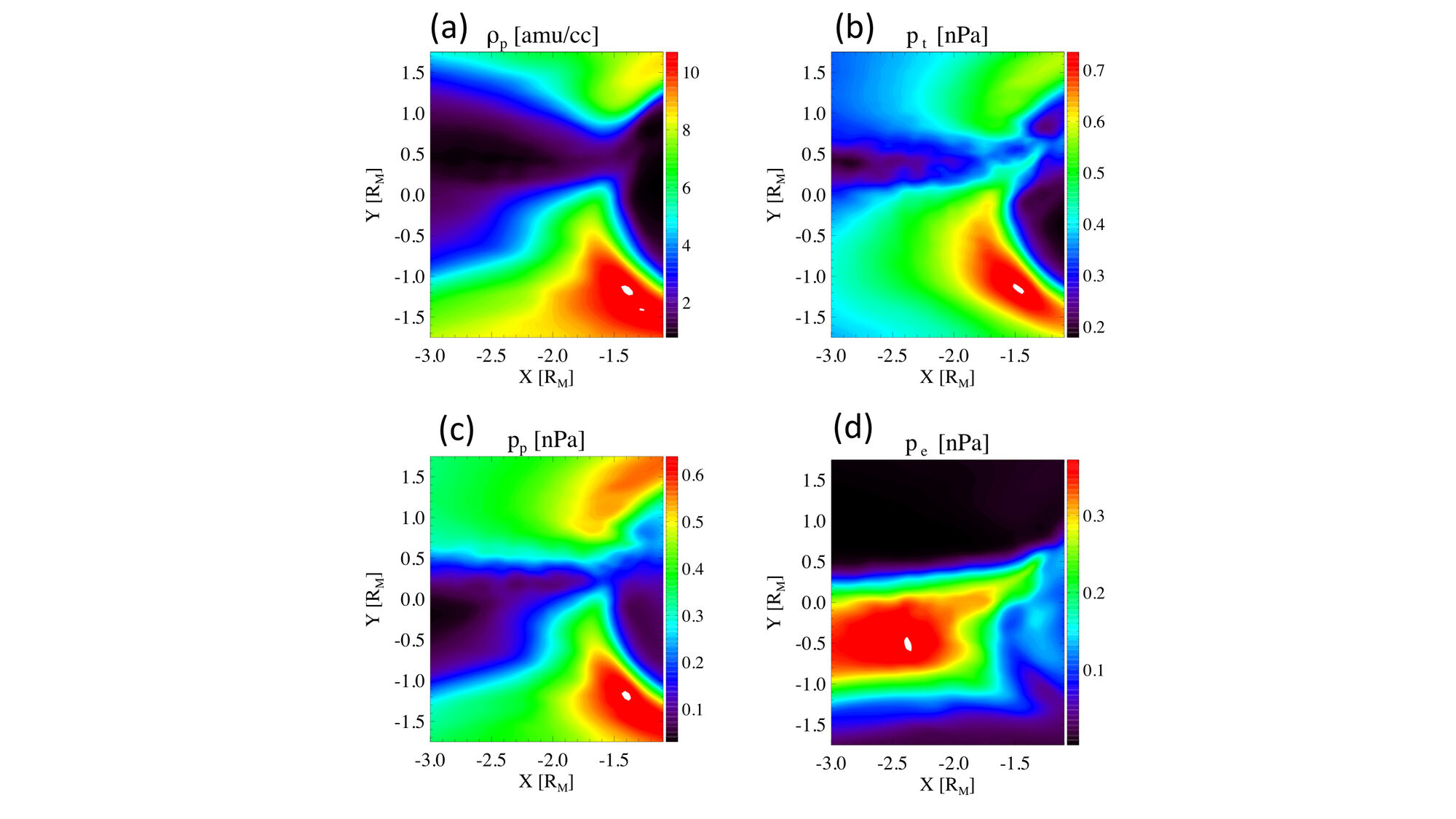}
 \caption{The time-averaged plasma profiles on the current sheet surface for Hall-A: proton density (a), total pressure (b), proton pressure (c) and electron pressure (d). The electron pressure
 presented here is calculated by a separate electron pressure equation in our MHD model.}
 \label{fig:cs_av_hall}
 \end{figure}

Figures~\ref{fig:cs_av_epic_a} to \ref{fig:cs_av_hall} show the time-averaged profiles of 
various plasma properties on the current sheet surface for MHD-EPIC-A, MHD-EPIC-B 
and Hall-A, respectively. The plots of Hall-B are not shown, because they are 
not significant difference than Hall-A in terms of the properties we discussed below. 
The current sheet surface is defined as the surface 
where $B_x$ changes sign, and its projection into the X-Y plane is shown in 
the figures. All three simulations show significant dawn-dusk asymmetries 
of plasma density, electron pressure and total pressure. In the inner magnetotail,
at a radial distance of $\sim 1.5\,R_M$ from the center of Mercury, the dawnside (-Y) plasma 
density (6~amu/cc for MHD-EPIC-B, and 10~amu/cc for MHD-EPIC-A/Hall-A) is 
about twice of the duskside (+Y) density (3~amu/cc for MHD-EPIC-B, 5~amu/cc for
MHD-EPIC-A, and 7~amu/cc for Hall-A). Both the density values and the dawn-dusk ratio are 
close to the  MESSENGER observation (Figure~\ref{fig:observation_plasma}). 
By studying Earth's magnetotail, \citet{Lin:2014} and \citet{Lu:2016} found 
that there is more plasma in the dawn sector of Earth's magnetotail due to 
the $\mathbf{E} \times \mathbf{B}$ drift caused by the 
Hall electric field. Since the Hall effect is the only reason to produce 
dawn-dusk asymmetry in the Hall-A simulation, Hall effect must be the 
reason to create higher dawnside plasma density in Hall-A as well as 
the MHD-EPIC simulations. The MESSENGER data indicates slight proton pressure 
enhancement on the dawnside (Figure~\ref{fig:observation_plasma}(b)), but our simulations 
do not show any significant preference of the proton pressure. The simulated electron 
pressure and hence the total pressure are higher on the dawnside (-Y). Eq.~(\ref{eq:pe}) 
is the electron pressure equation solved by the Hall-MHD model, and its right-hand side, 
the compression term, can produce the dawnside pressure enhancement. Because the $u_{e,z}$ 
component is small and the $u_{e,x}$ component changes slowly in the X-direction, the 
$\frac{\partial u_{e,y}}{\partial y}$ term must contributes most to the compression 
$\nabla \cdot \mathbf{u}_e$. From Figure~\ref{fig:current}(d), we can see the 
electron velocity reduces shapely from a few thousand to less than 500~km/s 
near $y=0.5\,R_M$. The braking of $u_{e,y}$ is consistent with the dawnside electron 
pressure enhancement. The amplitude of the proton velocity $u_{p,y}$ is much smaller 
than $u_{e,y}$, and so is the proton compression $\nabla \cdot \mathbf{u}_i$.
This may explain why there is no significant proton pressure asymmetry. Larger dawnside (-Y) 
electron pressure and total pressure are also consistent with thicker dawnside (-Y) 
current sheet thickness. 
 
\subsection{Magnetic reconnection}
\begin{figure}
 \includegraphics[width=1.0\textwidth, trim=0cm 0cm 0cm 0cm]{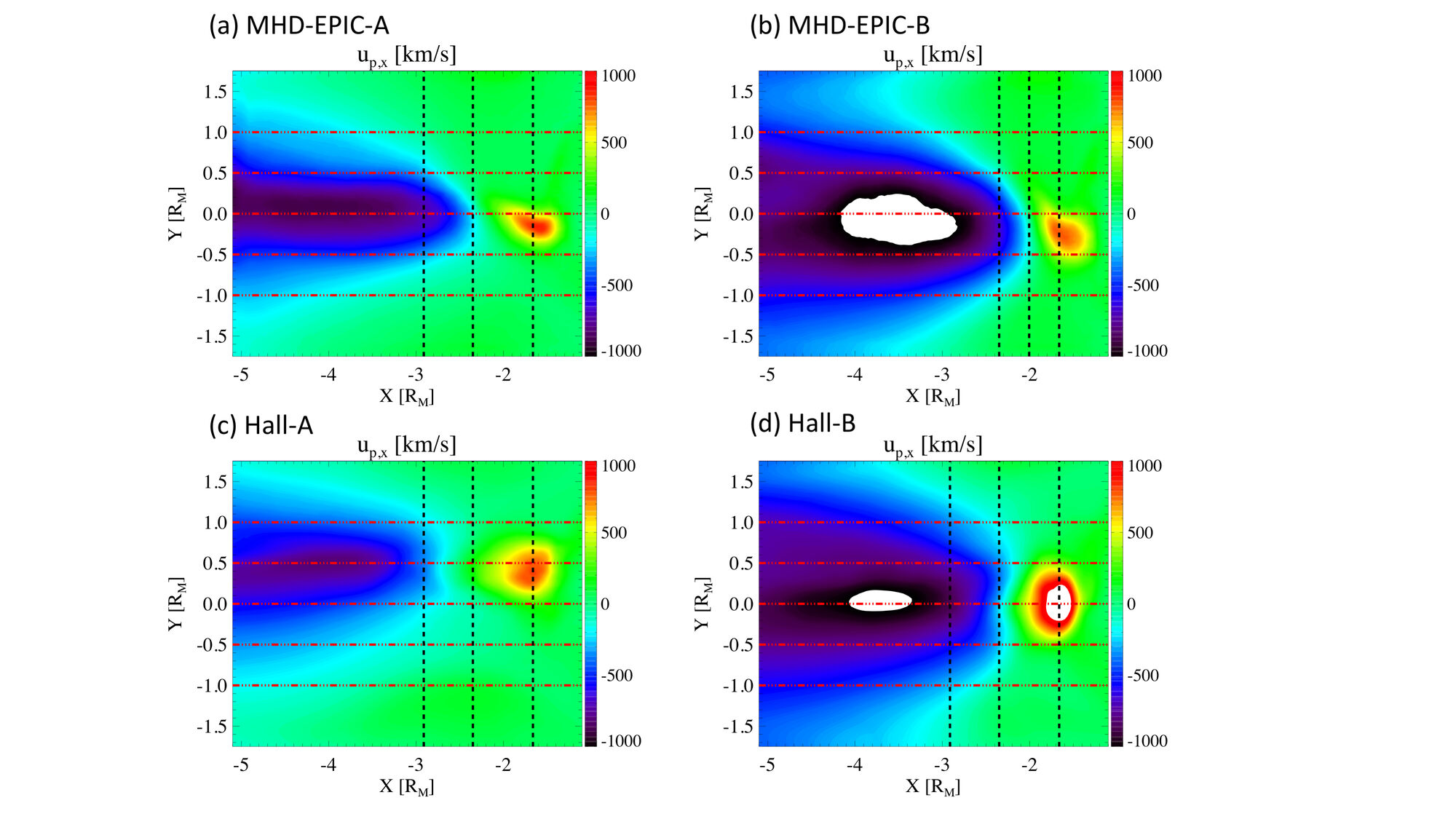}
 \caption{The time-averaged 
 X-component of the proton velocity on the current sheet surface for the four
 simulations. The horizontal 
 dashed lines are at $y=-1, -0.5, 0, +0.5$ and $+1\,R_M$, respectively.
 The location of the vertical black lines change from plot to plot. The time evolution along these 
 vertical lines are shown in the following figures. The color range is saturated at 1000 km/s and -1000 km/s.}
 \label{fig:cs_ux}
 \end{figure}

We discuss the asymmetries that are directly related to the magnetotail reconnection in
this section. The average proton reconnection jets on the current sheet surface are shown in Figure~\ref{fig:cs_ux}
for all simulations. In the MHD-EPIC simulations, there is no significant dawn-dusk asymmetry of 
the tailward jets. But it is clear that the planetward proton jets prefer the dawnside (-Y). 
In the Hall-A simulation, the reconnection jets center around $y=0.5\,R_M$, which is consistent with
the thin current sheet location (Figure~\ref{fig:cs_hall}). The Hall-B simulation does not show any 
significant dawn-dusk asymmetry of either tailward or planetward jets. 

\begin{figure}
 \includegraphics[width=1.0\textwidth, trim=0cm 0cm 0cm 0cm]{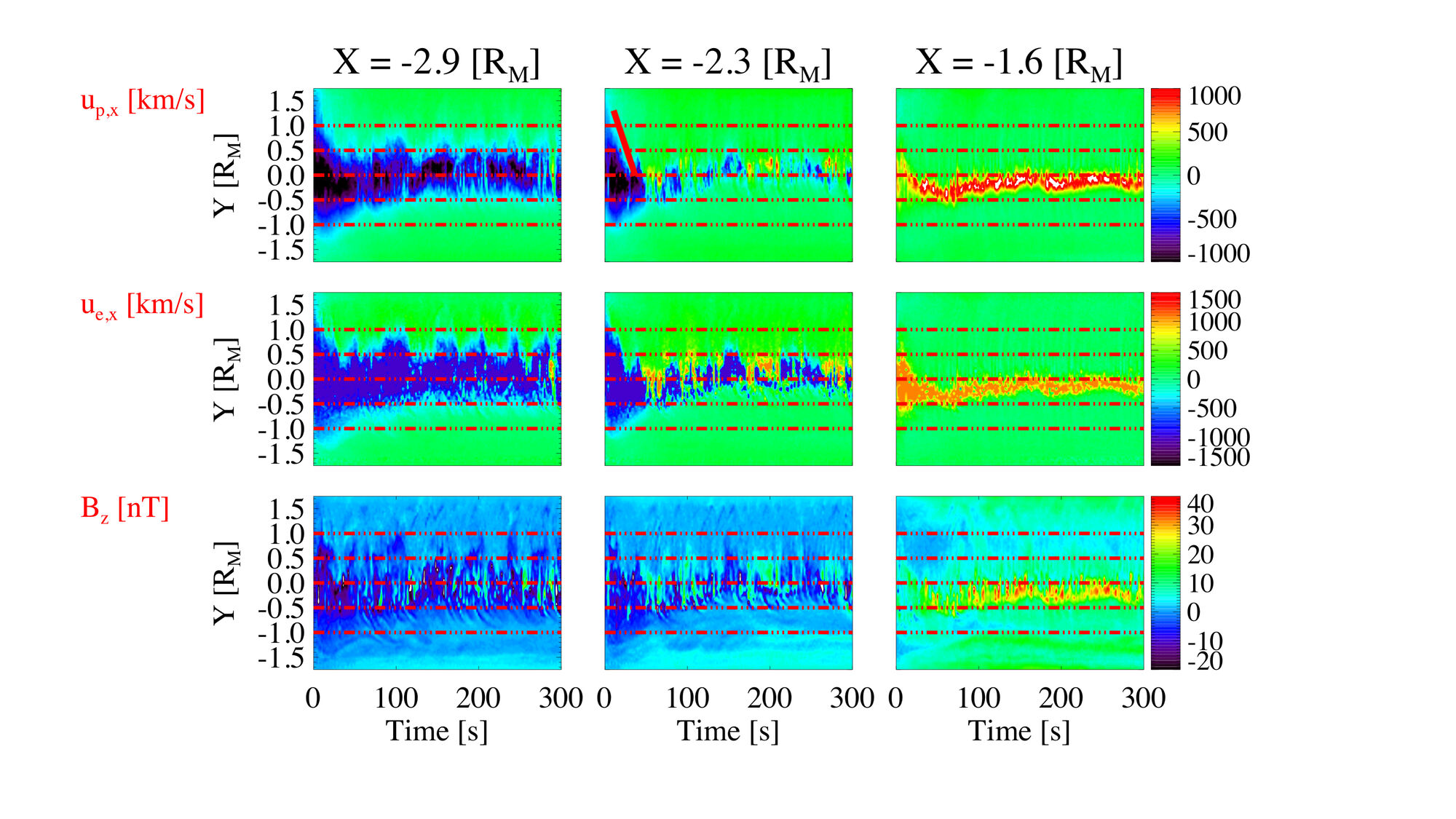}
 \caption{The evolution of different quantities on the current sheet surface at $x=-2.9\,R_M$, $x=-2.3\,R_M$ 
 and $x=-1.6\,R_M$ (see the three vertical lines in Figure~\ref{fig:cs_ux}(a)) for the MHD-EPIC-A simulation. 
 The time serials of 
 the x-component of the proton velocity $u_{p,x}$, the x-component of the electron velocity $u_{e,x}$, and the $B_z$ 
 magnetic field from the beginning of the simulation to the end are displayed.}
 \label{fig:time_epic_a}
 \end{figure}

The evolution of the proton jet $u_{p,x}$, electron jet $u_{e,x}$ and magnetic field $B_z$ in the 
current sheet center at $x=-2.9\,R_M$, $x=-2.3\,R_M$ and $x=-1.6\,R_M$ (the vertical 
lines in Figure~\ref{fig:cs_ux}(a)) are shown in Figure~\ref{fig:time_epic_a}. 
$x=-2.9\,R_M$ and $x=-1.6\,R_M$ are in the tailward and planetward outflow regions, respectively. 
$x=-2.3\,R_M$ is close to the X-lines so that the jets can be either tailward or planetward. 
If we ignore the first 50~s of the simulation, which corresponds to the transition period of 
starting MHD-EPIC from a steady-sate Hall MHD configuration, the reconnection sites and the tailward jets
shift to the dusk side slightly. For example, it is more frequent to observe electron jets
for $y\in [0,0.5]R_M$ than $y\in [-0.5,0]R_M$ at $x=-2.3\,R_M$. However, on the planet side of 
the X-line, both the high-speed plasma jets $u_{p,x}$ and $u_{e,x}$, and the enhanced $B_z$ shift
to the dawnside. At $x=-1.6\,R_M$, there are neither proton nor electron
jets found in the region $y>0$. 

\begin{figure}
 \includegraphics[width=1.0\textwidth, trim=0cm 0cm 0cm 0cm]{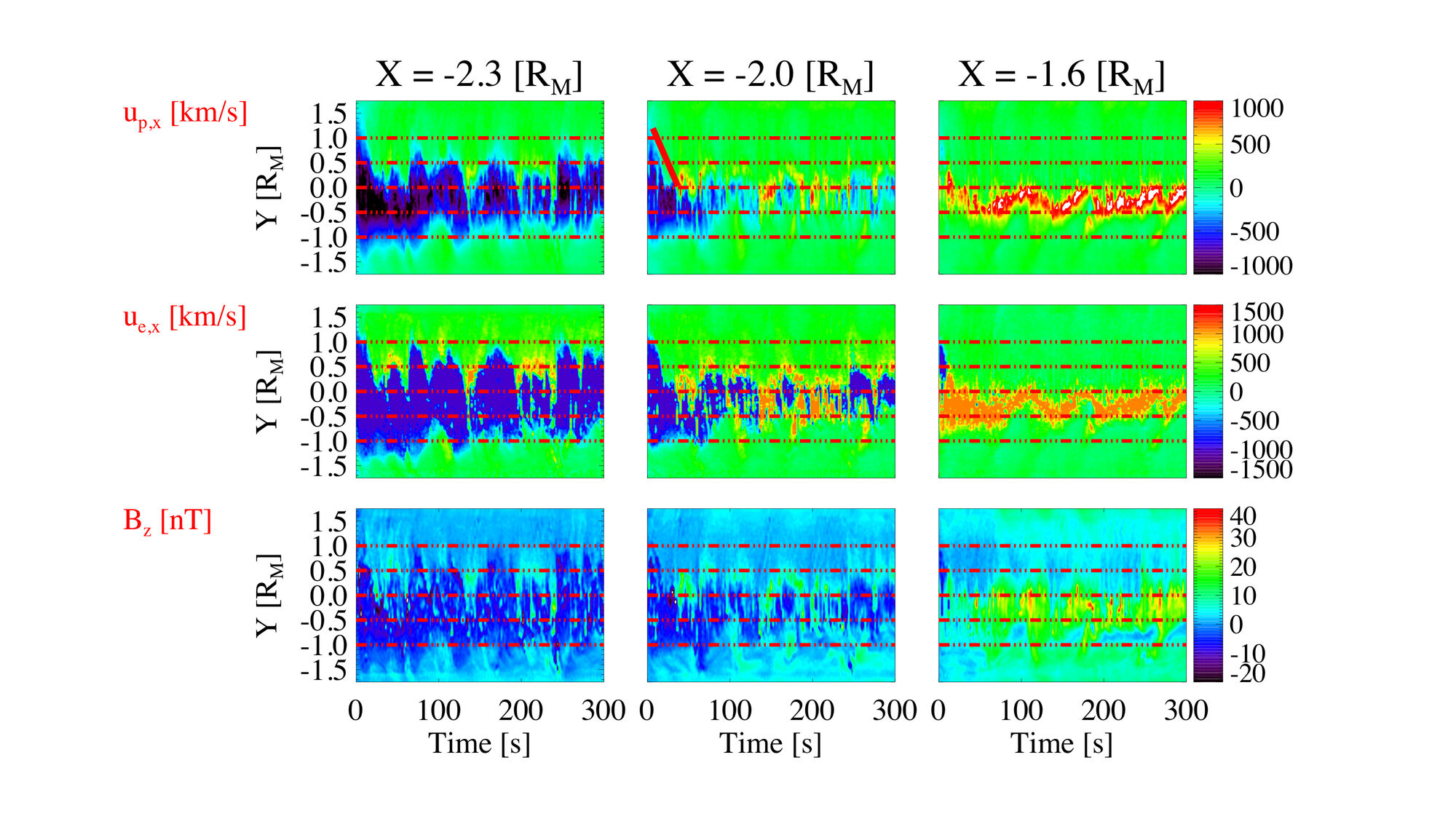}
 \caption{The same quantities as Figure~\ref{fig:time_epic_a} for the MHD-EPIC-B simulation at 
 $x=-2.3\,R_M$, $x=-2.0\,R_M$ and $x=-1.6\,R_M$ (see the three vertical lines in Figure~\ref{fig:cs_ux}(b)). }
 \label{fig:time_epic_b}
 \end{figure}

The reconnection products with a strong IMF driver (MHD-EPIC-B) are presented in 
Figure~\ref{fig:time_epic_b}. For this case, not only the planetward jets ($x=-1.6\,R_M$), but also 
the tailward jets ($x=-2.3\,R_M$) and the reconnection sites ($x=-2.0\,R_M$) shift to the
dawnside. For example, it is not unusual to see either proton jet $u_{p,x}$ or electron jet $u_{e,x}$
between $y=-0.5\,R_M$ and $y=-1.0\,R_M$ at $x=-2.3\,R_M$ and $x=-2.0\,R_M$, but it is rarer to 
have high-speed jets between $y=0.5\,R_M$ and $y=1.0\,R_M$ at the same X-coordinate. 

The simulated spatial distributions of the plasma jets and enhanced $B_z$ in the inner tail
are consistent with MESSENGER observations. Figure~2 of \citet{Poh:2017b} shows that the 
dawnside $B_z$ field is stronger than the duskside, and the $B_z$ field peaks at $y=-0.2\,R_M$. 
Our MHD-EPIC-A and MHD-EPIC-B simulations also show a peak value of $B_z \sim 30$~nT between
$y=0\,R_M$ and $y=-0.5\,R_M$ at $x=-1.6\,R_M$, and the dawnside $B_z$ is larger than the dusk side as well. 
\citet{Dewey:2017} found the energetic electron injections concentrate in the dawn sector, and the peak
fraction of the dipolarization associated events occurs at LT $\sim$ 1-2, which corresponds 
to $y \sim$ 0.4-0.9 for $x=-1.6\,R_M$. Our simulation results are consistent with the MESSENGER energetic particle observations. The simulation high-speed electron jets  
prefer to occur between $y=0\,R_M$ and $y=-0.5\,R_M$ at $x=-1.6\,R_M$. 

The MHD-EPIC simulations suggest that the closer to Mercury, the stronger the dawn-dusk 
asymmetries of the reconnection products are. Observational evidences for this pattern 
may already exist in the publications. \citet{Smith:2017} used an automated method to identify flux
ropes, and they observed a weak dawn-dusk asymmetry with $58\%$ of flux ropes observed in the dawn
sector. Most of the flux ropes lie between 1.5 and 2.5 $R_M$ down the tail. This statistical result
suggests that the dawn-dusk asymmetry between $x=-1.5\,R_M$ and $x=-2.5\,R_M$ is not very strong. 
But the energetic electron spatial distribution by \citet{Dewey:2017} shows that almost all injections
are observed in the midnight-to-dawn sector. Even though these two papers discussed different phenomena, both
phenomena are likely the products of magnetic reconnection. In order to further confirm this hypothesis,
we plot the spatial distribution of the dipolarization fronts observed by MESSENGER 
in Figure~\ref{fig:observation_df}, which shows strong dawn-dusk asymmetry, and there is
a trend that the asymmetry is stronger in the region closer to Mercury. 
Figure~\ref{fig:df} shows the evolution of a dipolarization event, which is characterized by 
$B_z$ enhancement, from the MHD-EPIC-A simulation. The structure of enhanced $B_z$ is circled 
by the red ovals on the plots. The dipolarization initially appears at $x\sim -2.3\,R_M$, and 
the majority of the structure
is in the dusk sector. The enhanced $B_z$ structure moves dawnward when it is moving towards
Mercury. The electron flow streamlines are over-plotted above $B_z$. It is clear that
the electrons move in the same direction as the dipolarization front. The dawnward velocity
component of electrons is a natural consequence of the cross-tail current. If we assume that 
part of the dawnward moving electrons are frozen into the magnetic field lines, the motion
of the dipolarization front can be explained as well. The protons around the dipolarization front
are moving duskward in the current sheet (see the proton streamlines in Figure~\ref{fig:df}). 
However, the high-speed proton jet still prefers the dawnside in our simulations, which is consistent with 
MESSENGER observations \citep{Sun:2017}. The difference between the proton
motion direction and the high-speed jet preferential direction suggests that the fast proton
flows observed far away from the reconnection sites are not direct products of the magnetic 
reconnection itself. Instead, these fast protons may be accelerated by the dipolarization 
fronts \citep{Zhou:2010}.

\begin{figure}
 \includegraphics[width=1.0\textwidth, trim=0cm 0cm 0cm 0cm]{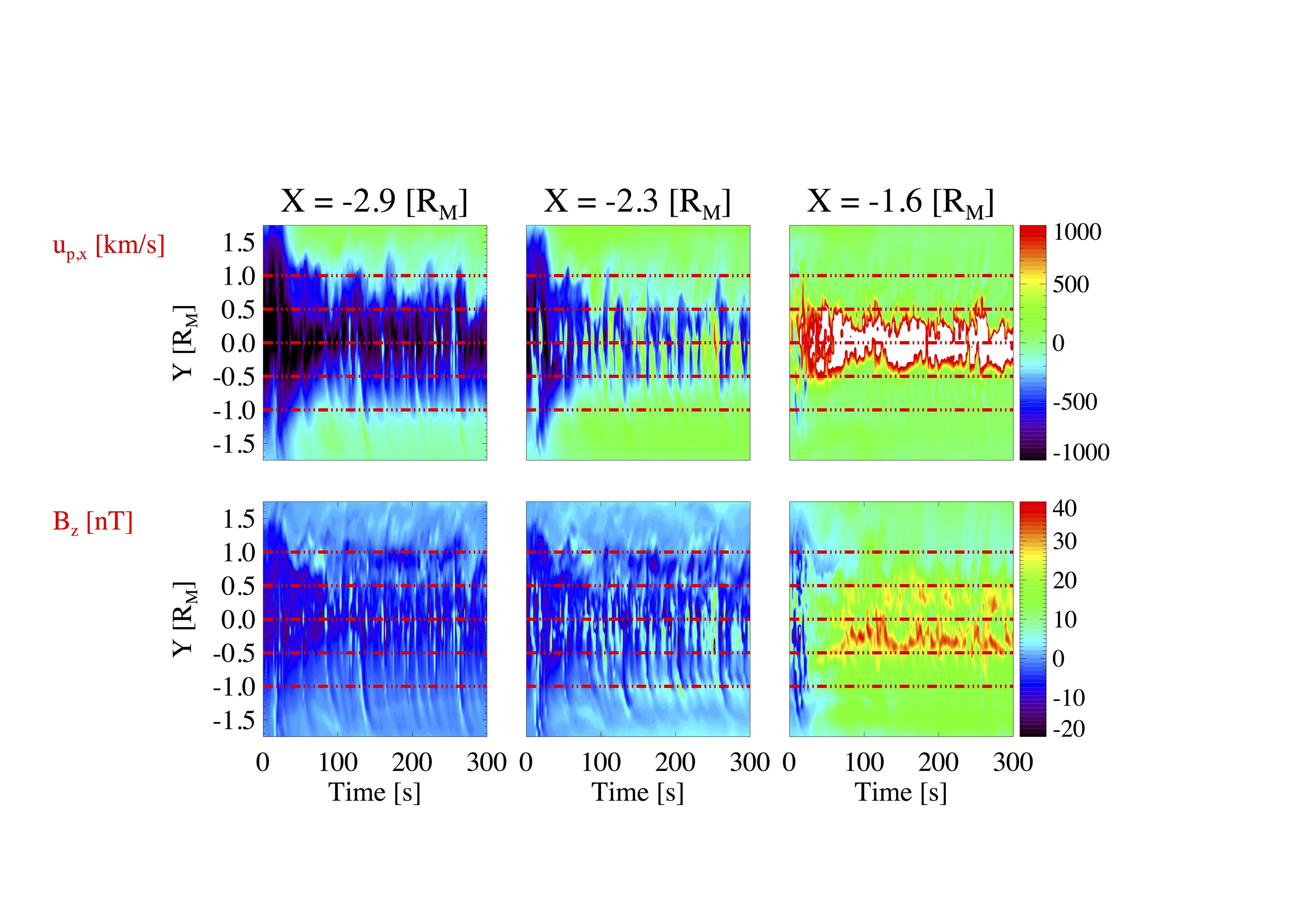}
 \caption{The evolution of the proton velocity $u_{p,x}$ and magnetic field $B_z$ at  
 $x=-2.9\,R_M$, $x=-2.3\,R_M$ and $x=-1.6\,R_M$ (see the three vertical lines in Figure~\ref{fig:cs_ux}(d)) 
 of the current sheet of the Hall-B simulation.  }
 \label{fig:time_hall_b}
 \end{figure}

\begin{figure}
 \includegraphics[width=1\textwidth, trim=0cm 0cm 0cm 0cm]{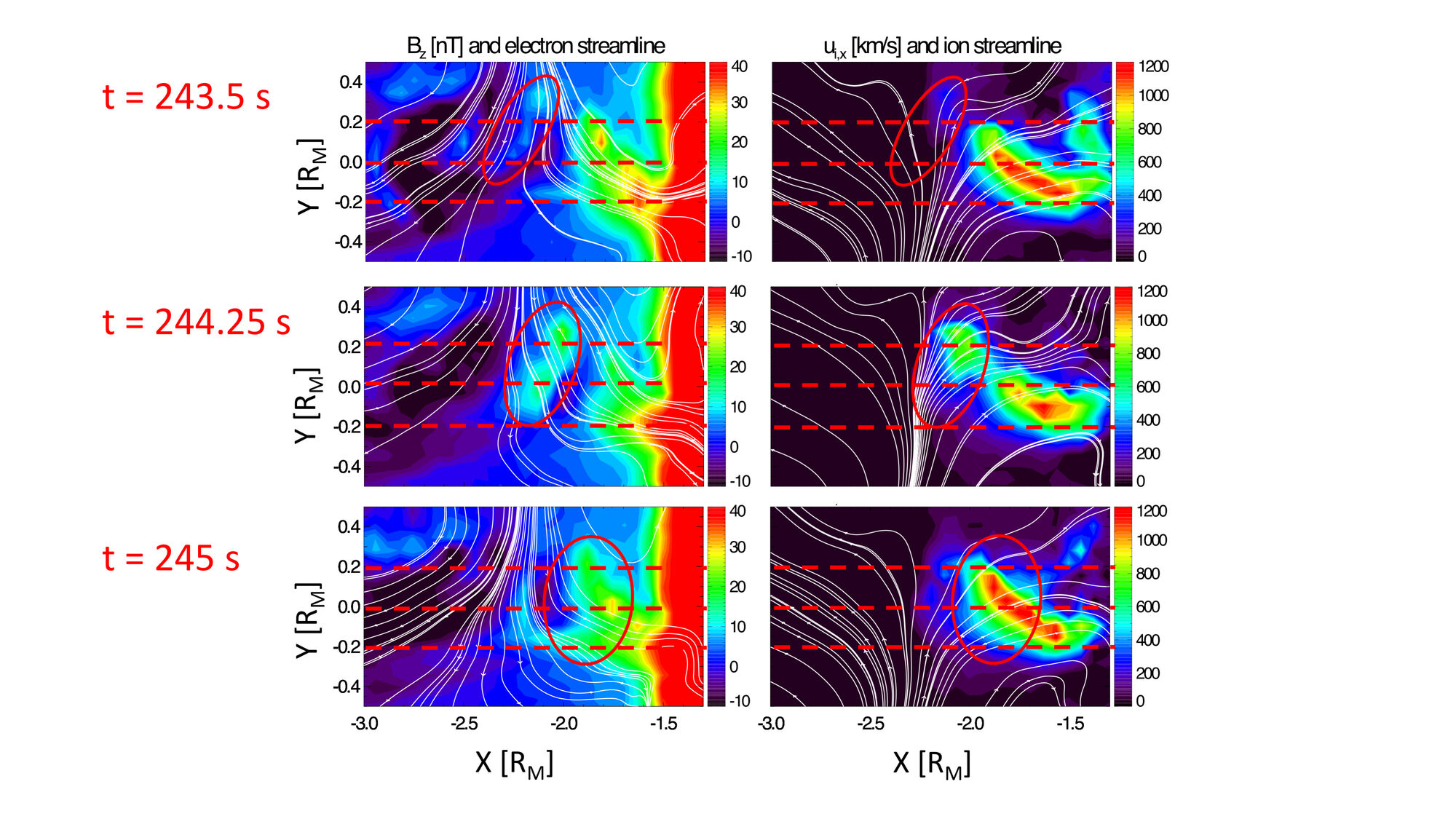}
 \caption{The $B_z$ magnetic field and proton velocity $u_{p,x}$ on the current sheet surface at different times. 
 The electron streamlines (the white lines) are overplotted on the $B_z$ plots. The red ovals indicate the 
 location of enhanced $B_z$.}
 \label{fig:df}
 \end{figure}

In order to demonstrate the importance of including the kinetic effects into the model, we compare
the MHD-EPIC simulations with pure Hall-MHD simulations. Figure~\ref{fig:time_hall_b}
shows the evolution of plasma jets and $B_z$ for Hall-B simulation. This simulation does not show
any significant dawn-dusk asymmetry and the results are quite different from those of the MHD-EPIC-B run.


\section{Discussion} 
\label{section:discussion}
  \begin{figure}
 \includegraphics[width=0.9\textwidth, trim=0cm 0cm 0cm 0cm]{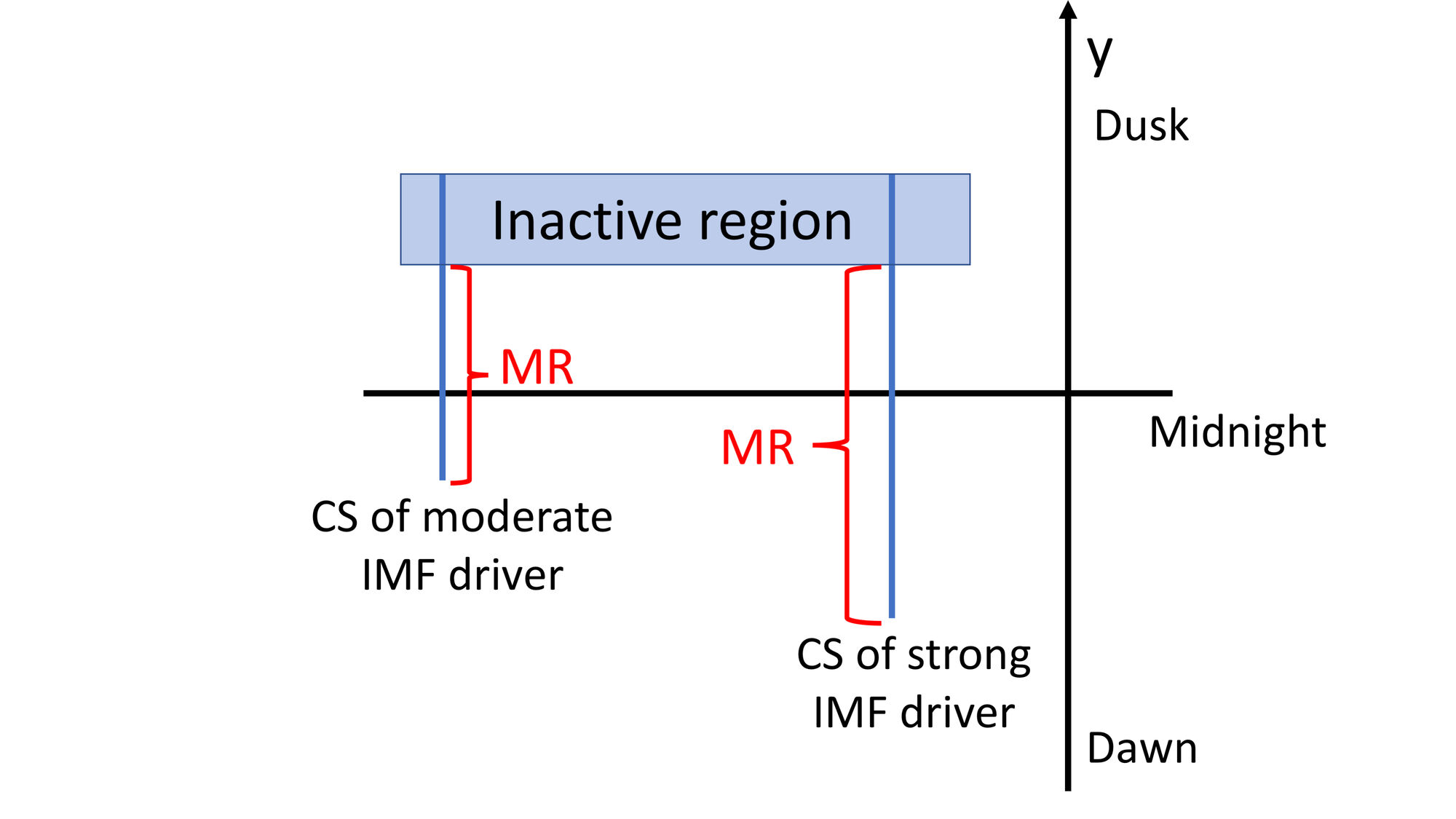}
 \caption{A cartoon illustrating the influence of the current sheet asymmetry and reconnection 
 inactive region on the reconnection asymmetry for moderate and strong IMF driving conditions. The magnetic reconnection occurs in the region indicated with MR.}
 \label{fig:cartoon}
 \end{figure} 
 
In the MHD-EPIC-A simulation, the IMF driver of $B_z=-8.5$\,nT is moderate. The 
driver of MHD-EPIC-B is strong. These simulations suggest that Mercury's 
magnetotail reconnection sites slightly shift to the duskside (Figure~\ref{fig:time_epic_a}) 
when the dawnside current sheet is significantly thicker than the duskside 
(Figure~\ref{fig:cs}(a) and (b)) under a moderate IMF driver (MHD-EPIC-A), and the
reconnection sites shift to the dawnside (Figure~\ref{fig:time_epic_b}) when the 
dawnside current sheet is almost as thin as the duskside (Figure~\ref{fig:cs}(c) 
and (d)) under a strong driver (MHD-EPIC-B). The results of MHD-EPIC-B simulation are consistent
with what \citet{Liu:2019} found in 3D box PIC simulations. They found that 
there is a reconnection `inactive' region on the ion drifting side (the 
duskside in our simulations) of a thin current sheet, so that the magnetic reconnection prefers
the electron-drifting side. Under a moderate driver, the majority of the thin current sheet 
lies on the duskside. For such current sheet configuration, even though part of the duskside
current sheet is inactive, most of the reconnection sites may still be on the duskside, 
just as in the MHD-EPIC-A simulation. Since part of the duskside current sheet is inactive, the 
duskside preference of the reconnection should be weaker than the thin current sheet. 
We think this may be the reason why the MHD-EPIC-A simulation
shows strong current sheet thickness asymmetry, but the reconnection preference is not significant.
When the IMF driver is strong, such as in the case of MHD-EPIC-B, the current sheet is thin enough to allow 
magnetic reconnection to occur in almost the whole magnetotail current sheet, so that the dawn-dusk
asymmetry of the current sheet thickness has little influence on the magnetic reconnection, and the
inactive region \citep{Liu:2019} on the duskside determines the dawn-dusk asymmetry of the reconnection sites. 
Figure~\ref{fig:cartoon} displays the relative importance of the current sheet asymmetry
and the reconnection inactive region. 
Besides the dawnside preference introduced by the `inactive' region, we find that 
the planetward moving electron jets and the dipolarization fronts are also shifted dawnward. The dawnward
motion  makes it rare to observe high-speed planetward plasma jets and dipolarization events in the dusk sector. 

We now turn to the first 50 s of the MHD-EPIC simulations. Since it corresponds to the transition
from the steady-state Hall-MHD to MHD-EPIC, the results of the first 50 s may not represent a typical 
state of Mercury's magnetotail. But it still provides interesting insights into Mercury's magnetotail 
reconnection. At the very beginning, MHD-EPIC inherits the current sheet structure from the steady-state
Hall-MHD results. The Hall effect of the steady-state Hall-MHD exists, but it is weak 
due to the large numerical diffusion. The current sheet thickness between $y=-1.0\,R_M$ 
and $y=1.0\,R_M$ is less than $0.2\,R_M$ and is approximately symmetric. The X-lines estimated from 
the tailward jets (Figure~\ref{fig:time_epic_a} and Figure~\ref{fig:cs_av_epic_b}) are more 
than $1\,R_M$ wide in the cross-tail direction initially. As soon as the MHD-EPIC simulation starts, the duskside X-lines start to shrink (solid red lines in Figure~\ref{fig:time_epic_a} and Figure~\ref{fig:cs_av_epic_b}), 
so that almost all the reconnection sites are in the dawn sector at t = 30 s. The shrinkage 
of the X-lines may be related to the reconnection inactive region discussed by \citet{Liu:2019}.  

The MESSENGER observations of current sheet thickness \citep{Poh:2017b}, flux ropes,
dipolarization events \citep{Sun:2016, Sun:2017, Smith:2017} and energetic electron 
events \citep{Dewey:2017} do not and cannot distinguish the events under different 
IMF conditions. For the current sheet thickness observation, the 
current sheet sampling is almost uniform in time. If the moderate IMF condition dominates 
throughout the period during which the MESSENGER observations were obtained, the 
asymmetric current sheet (like MHD-EPIC-A) will contribute most sampling data 
points in the statistics. However, strong IMF driving is likely to produce magnetotail 
reconnection products more frequently. Even if the moderate IMF condition occurs 
more frequently, it is still possible that most observed reconnection related events 
are produced by strong IMF drivers. 

\begin{figure}
 \includegraphics[width=\textwidth]{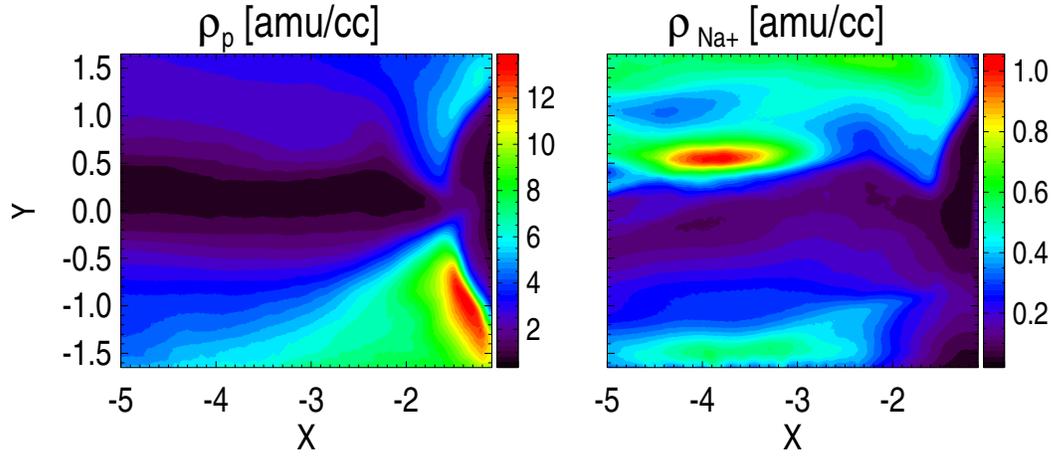}
 \caption{The proton and sodium density on the current sheet surface for multispecies-MHD-EPIC.}
 \label{fig:na}
 \end{figure}

Our model assumes all the ions are protons.  The heavy ions, such as sodium, have not been incorporated 
into the simulations. The model does not produce Kelvin-Helmholtz instability (KHI) on 
either side of the magnetopause. But our MHD-EPIC simulations still manifest the dawn-dusk asymmetries 
that are comparable with observations, which suggests that neither heavy ions nor KHI are necessary for the 
reconnection related dawn-dusk asymmetries, even though they may still play an important role. We have 
tried to incorporate sodium into our MHD model by using multispecies MHD, and therefore the sodium will 
also be treated as a separate ion species inside the PIC region \citep{Ma:2018}. The sodium ions
enter the simulation domain from the MHD inner boundary. To be 
specific, we set the sodium mass density to be $70\%$ of the total mass density in the inner 
boundary ghost cells. This mass density matches a number density of $\sim 10\%$, which is the heavy ion
abundance in the plasma sheet observe by MESSENGER \citep{Gershman:2014}. The  boundary condition 
does not introduce any dawn-dusk asymmetry by itself. This preliminary simulation shows the duskside 
sodium density is indeed higher than the dawnside in the current sheet (Figure~\ref{fig:na}), which is
consistent with MESSENGER observations. This simulation does not show any significant difference 
compared to the one with single ion species. Our current inner boundary condition relies 
on numerical diffusion to get sodium into the simulation domain from Mercury's surface and 
the sodium density inside the current sheet  is still lower than observed by MESSENGER 
 \citep{Gershman:2014}, so we cannot draw any conclusion about the role of
heavy ions so far. We will explore the role of heavy ions with an improved model in the future. 

The MHD-EPIC-B simulation demonstrates that magnetic reconnection prefers the dawnside, and both MHD-EPIC-A 
and MHD-EPIC-B show the planetward high-speed plasma flows and dipolarization events move toward the dawnside. 
But it is still rare to see tailward jets beyond $y=-1.0\,R_M$ or to see planetward jets beyond  $y=-0.5\,R_M$. MESSENGER
observed many such events far away from the midnight direction, such as the dipolarization fronts in 
Figure~\ref{fig:observation_df} and statistics from other papers \citep{Sun:2016, Smith:2017, Dewey:2017}.  
This discrepancy may be simply caused by the varying IMF in the observations. It can also be introduced by 
the physics that is not in our model, such as a proper heavy ion profile.

Both the MHD-EPIC and pure Hall-MHD simulations presented in this study show that the duskside current 
sheet is thinner than the dawnside, but the thickness obtained from Hall-MHD is significantly larger 
than that of MESSENGER observations and MHD-EPIC simulations. Magnetic reconnection of Hall-A 
simulation shifts to the duskside significantly. There isn't any significant dawn-dusk asymmetry 
in the Hall-B simulation. In general, Hall-MHD simulations do not appear to match observations 
very well in terms of dawn-dusk asymmetries of magnetic reconnection.

\section{Summary}
We use the MHD-EPIC model to study dawn-dusk asymmetries of Mercury's magnetotail. The simulation results, 
such as the current sheet thickness, plasma density asymmetry, and reconnection asymmetry, agree with MESSENGER
observations. The key simulation results are:
\begin{itemize}
\item The dawnside plasma density and electron pressure are higher than the duskside. The proton pressure 
does not exhibit significant dawn-dusk asymmetry in the simulations.  
\item The dawnside current sheet is thicker than the duskside. 
\item When the IMF driver is moderate, for example, $B_z = -8.5\,nT$, the current sheet 
thickness asymmetry is strong, and 
the magnetotail X-lines may shift to the duskside. When the IMF driver is strong, for 
example, $B_z=-19.4\,nT$, the current sheet thickness asymmetry is not significant, and the 
magnetotail reconnection prefers the dawnside. 
\item  The dipolarization events and the planetward high-speed plasma flows, including both 
proton flows and electron flows, concentrate in the dawn sector.
\item The preliminary multispecies-MHD-EPIC simulation produces higher duskside sodium density 
in the current sheet but does not change the asymmetry of the reconnection significantly. 
\end{itemize}

\acknowledgments
This work was supported by the INSPIRE NSF grant PHY-1513379, the NSF                     
PREEVENTS grant 1663800, the Impacts of Extreme Space Weather Events 
on Power Grid Infrastructure project funded by the U.S. Department of 
Energy (DE-AC52-06NA25396) through the Los Alamos National Laboratory 
Directed Research and Development program, the NASA Living With a Star program 
grant NNX16AJ67G and the NASA's Solar SystemWorkings program grant NNX15AH28G.
Computational resources supporting this work were                   
provided on the Blue Waters super computer by the NSF PRAC grant ACI-1640510,                
on the Pleiades computer by NASA High-End Computing (HEC) Program through                    
the NASA Advanced Supercomputing (NAS) Division at Ames Research Center,                     
and from Cheyenne (doi:10.5065/D6RX99HX) provided by NCAR's Computational and              
Information Systems Laboratory, sponsored by the National Science Foundation.

MESSENGER data used in this study were available from the Planetary Data
System (PDS): http://pds.jpl.nasa.gov. The MESSENGER project is supported by
the NASA Discovery Program under contracts NASW-00002 to the Carnegie
Institution of Washington and NAS5-97271 to The Johns Hopkins University
Applied Physics Laboratory.

The SWMF code (including BATS-R-US and iPIC3D) is publicly available through the 
csem.engin.umich.edu/tools/swmf web site after registration. 
The output of the simulations presented in this paper can be obtained 
by contacting the first author Yuxi Chen.

\clearpage                                                                                          
\bibliography{csem,yuxichen}   

\end{document}